\documentclass[twoside]{article}

\usepackage[accepted]{aistats2021}
\usepackage[round]{natbib}

\usepackage{times}
\usepackage{latexsym}
\usepackage{amsmath}
\usepackage{lipsum}
\usepackage{amssymb}
\usepackage{float}
\usepackage{graphicx}

\usepackage{algorithm}
\usepackage{algpseudocode}
\usepackage{sidecap}

\newtheorem{theorem}{Theorem}
\newtheorem{definition}{Definition}

\begin{document}



\twocolumn[

\aistatstitle{\emph{Surprise}: Result List Truncation via Extreme Value Theory}

\aistatsauthor{ Dara Bahri \And Che Zheng \And Yi Tay \And Donald Metzler \And Andrew Tomkins}

\aistatsaddress{Google Research \And Google Research \And Google Research \And Google Research \And Google Research}
]

\begin{abstract}
Work in information retrieval has largely been centered around ranking and relevance: given a query, return some number of results ordered by relevance to the user. The problem of \textit{result list truncation}, or where to truncate the ranked list of results, however, has received less attention despite being crucial in a variety of applications. Such truncation is a balancing act between the overall relevance, or usefulness of the results, with the user cost of processing more results. Result list truncation can be challenging because relevance scores are often not well-calibrated. This is particularly true in large-scale IR systems where documents and queries are embedded in the same metric space and a query's nearest document neighbors are returned during inference. Here, relevance is inversely proportional to the distance between the query and candidate document, but what distance constitutes relevance varies from query to query and changes dynamically as more documents are added to the index. In this work, we propose \textit{Surprise} scoring, a statistical method that leverages the Generalized Pareto Distribution that arises in Extreme Value Theory to produce interpretable and calibrated relevance scores at query time using nothing more than the ranked scores. We demonstrate its effectiveness on the result list truncation task across image, text, and IR datasets and compare it to both classical and recent baselines. We draw connections to hypothesis testing and $p$-values.
\end{abstract}

\section{Introduction}

Work in information retrieval systems has traditionally focused on improving ranking while the problem of determining \textit{how many} results to return has received less attention. This task, referred to as \textit{result list truncation}, is defined as follows: for a specific query and its ranked result list, determine the number of candidates that should be returned such that some evaluation metric, like F1, is optimized. If the relevance scores are well-calibrated across queries, then one effective strategy is to pick a \textit{single} global decision threshold on the scores. If they are not, then a parametric model can be learned to find the optimal cutoff conditioned on the query. As we will see however, in common IR systems based on ``metric learning'', this latter approach is challenging or impossible.

In this work we propose a statistical method called \textit{Surprise} scoring for calibrating a single result list by capturing how surprising or unexpected each candidate score is under the null hypothesis that it comes from a non-relevance distribution we fit. Our method remedies many of the concerns of existing approaches and is broadly applicable. It is premised on a single condition -- that the IR system has good ranking performance (i.e. it scores relevant results over non-relevant ones with high probability) and that, naturally, the system returns the best results it has for a particular query.

\begin{figure}[!ht]
\centering
\begin{tabular}{cc}
Train & Test \\
\includegraphics[width=0.215\textwidth]{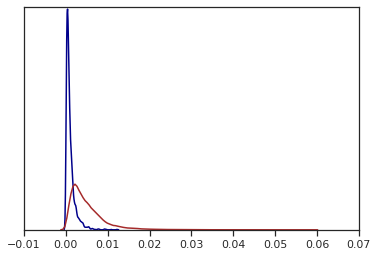}  & \includegraphics[width=0.215\textwidth]{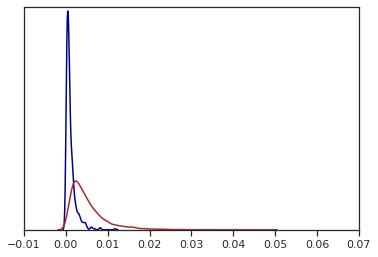} \\
\multicolumn{2}{c}{Same Document (negative cosine)} \\
\includegraphics[width=0.215\textwidth]{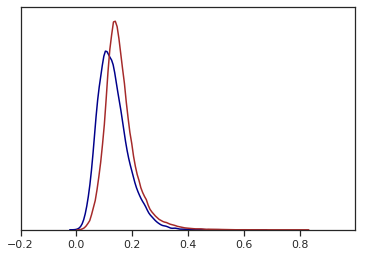}  & \includegraphics[width=0.215\textwidth]{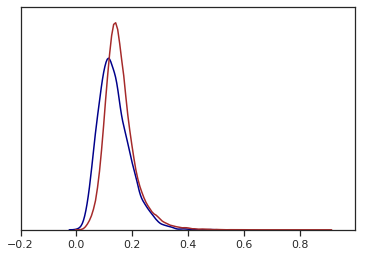} \\
\multicolumn{2}{c}{Omniglot (euclidean)} \\
\includegraphics[width=0.215\textwidth]{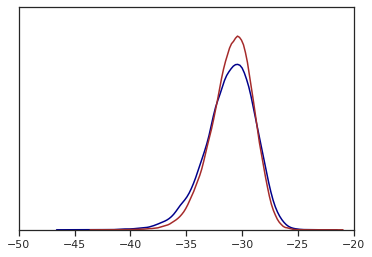}  & \includegraphics[width=0.215\textwidth]{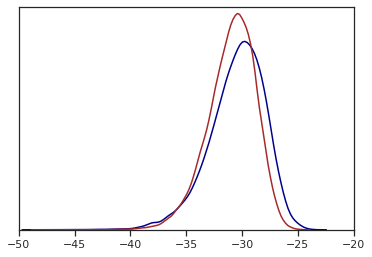} \\
\multicolumn{2}{c}{News Authorship (negative dot product)} \\
\end{tabular}

\caption{Distance distributions across queries for different datasets. Blue distributions capture distance to the first relevant sample while red distributions capture distance to the closest 10 non-relevant samples. Cosine and dot product similarities are negated to ensure lower means closer. When aggregated together, it is hard to distinguish between relevant and non-relevant results from the distances alone. This is a result of distances being uncalibrated across the embedding space.}
\label{fig:distance_score_dist}
\end{figure}

In metric learning, an encoder model is machine-learned to embed documents and queries in the same metric space such that the most relevant documents for a query are its nearest neighbors in this space. Typically, the encoder is trained using (query, document) pairs, where positive examples comprise queries with their relevant documents and negative ones comprise queries with a random sample of non-relevant documents. The loss function, usually contrastive or triplet loss, encourages positive pairs to be closer together than negative ones. After learning an encoder, an input dataset is encoded and indexed in a scalable retrieval system that supports efficient, exact or approximate nearest neighbor lookup. During serving, the query's $k$ nearest results in the embedding space are fetched and returned in ranked order of closeness, where relevance is inversely proportional to the distance between the query and result. Unfortunately however, as shown in Figure~\ref{fig:distance_score_dist}, what distance constitutes relevance is highly query dependent and so choosing a single global distance threshold to separate relevant and non-relevant results is prone to error. Existing losses only care about ranking order or relative distances within a mini-batch of examples and achieving calibrated distance scores across all queries involves joint optimization over \textit{all} examples, which is quite often infeasible. Moreover, the distance distributions are constantly changing as documents are added to the index. These reasons make list truncation particularly challenging in this setting.

To this end, we present Surprise scoring, a broadly applicable method for constructing calibrated scores using nothing more than a single ranked list of raw distance or relevance scores.
The main contributions of this paper are:
\begin{enumerate}
\item We present a statistically-grounded method for rescoring a ranked list of relevance or distance-based scores. Our method consists of (1) a novel rescoring function and (2) a new greedy method for selecting its hyper-parameters. Unlike many other parametric methods, the rescoring is learned on a \textit{single} example and is thus particularly useful in applications where training a parametric model on all examples is challenging due to, for example, dataset size or frequent distribution shifts. Furthermore, much like the concept of $p$-values in hypothesis testing, our scores are interpretable and well-calibrated across settings.
\item We apply our proposed method to the result list truncation task in two distinct settings over four datasets:
\begin{itemize}
\item When queries and results are represented in an embedding space and retrieval is efficient nearest-neighbor search that fetches a small neighborhood of the query from a massive index. In this case, we leverage only the \textit{local} neighborhood to select a cutoff threshold for truncating the ordered list of rescored candidates. This large-scale setting is becoming increasingly prevalent in the IR community.
\item In the more general setting where the indexing and retrieval system can take any form. In this case, our rescoring algorithm remains the same, but we leverage \textit{global} training data to pick the cutoff threshold for the rescored candidates.
\end{itemize}
We show that our proposed method often outperforms both classical and recent competitive baselines for result list truncation.
\end{enumerate}

\section{Related Work}\label{sec:related}
\subsection{Modeling Relevance Scores}
There has been extensive research focused on understanding and modeling score distributions of retrieval systems. Most of the early work in this area focused on fitting parametric probability distributions to score distributions \cite{manmathaSIGIR01,arampatzisSIGIR09}. This is often accomplished by assuming that score distributions can be represented as a mixture of a relevance distribution and a non-relevance (or background or noise) distribution. The parameters of these distributions are typically fit for a given system using the expectation-maximization (EM) algorithm.

More recently, there has been research into how to leverage machine learning to  optimally truncate a ranked list with respect to some metric of interest. For example, in cascade-style ranking systems \citep{wangSIGIR11} the goal is balance between effectiveness and efficiency. Recent work investigated a number of machine learning approaches for dynamically determining cutoffs within cascade-style ranking systems \citep{culpepperADCS16}. In addition, the Transformer and Bi-directional Long Short-Term Memory (LSTM) neural architectures have been utilized to identify the best position to truncate a given ranked list \citep{bahri2020choppy,lienICTIR19}.

The work presented here is also closely related to the task of query performance prediction \citep{cronentownsendSIGIR02}. The goal of this task is to automatically determine the effectiveness of a given query. If there existed a system that could perfectly predict the effectiveness of a query with respect to a set of results, it could be leveraged to determine the best set of results to the user for any given effectiveness measure. Approaches to query performance prediction include pre-retrieval approaches \citep{hauffCIKM08}, relevance modeling-based approaches \citep{cronentownsendSIGIR02,zhouSIGIR07}, and more recently neural network-based approaches \citep{zamaniSIGIR18}.

Determining how many results to return to users arises in a number of practical applications. For example, in sponsored search, displaying too many irrelevant ads to users may frustrate them and result in so-called ``query blindness''. As a result, there has been research that investigated whether it is possible to determine whether any ads should be returned to the user or not \citep{broderCIKM08}. A similar problem formulation investigated how many ads should be returned to the users \citep{wangPAKDD11}. Determining the optimal number of results to return is also important in a number of other search tasks, including legal e-discovery  \citep{tomlinsonTREC07}, where there is an immense cost associated with reviewing results.

The ability to effectively calibrate scores across queries and corpora has also been studied in the context of federated search tasks \citep{shokouhiFTIR11}, such as meta-search \citep{montagueCIKM01}.

\subsection{Extreme Value Theory}
Extreme value theory (EVT)~\citep {pickands1975statistical}, also referred to as extreme value analysis, is a subfield of statistics dealing with extreme deviations from the mean of probability distributions. More precisely, given an ordered sample of some random variable, it attempts to estimate the probability of events that are more extreme than any observed earlier. Usage of EVT spans many disciplines, including finance / risk management, traffic prediction, geological engineering, structural engineering, and earth sciences. Some concrete examples include modeling extreme floods and the magnitude of freak waves, financial market risk, mutational events during evolution, and large wildfires ~\citep{abarbanel1992statistics,alvarado1998modeling,castillo2012extreme,embrechts2013modelling}.

Within the context of machine learning, Extreme Value Machines~\citep{rudd2017extreme} were proposed, which are able to perform incremental learning in the presence of unknown query classes. \cite{vignotto2020extreme} suggest techniques for anomaly detection that improve the Extreme Value Machine.
Meanwhile, \cite{weng2018evaluating} leverage EVT in ``Cross Lipschitz Extreme Value for nEtwork Robustness'', a new measure of a neural network's robustness to adversarial examples. EVT has also been used in applications like learning to rank~\citep{zong2014learning}.

\section{Preliminaries}\label{sec:preliminaries}
We begin by reviewing the statistical underpinnings of our method.  

\begin{definition}[Conditional Excess Distribution]
The conditional excess distribution of a random variable X is $F_u$, where 
\begin{align*}
  F_{u}(x) = P(X - u \leq x | X > u). 
\end{align*}
\end{definition}

\begin{definition}[Maximum Domain of Attraction]
Let $\xrightarrow{dist.}$ denote convergence in distribution. A distribution F belongs to the Maximum Domain of Attraction (MDA) of a Generalized Extreme Value distribution (i.e. Fr\'{e}chet, Weibull, Gumbel) H if and only if there exists constants $c_n > 0$ and $d_n$ such that if $X_i \sim F$,
\begin{align*}
\frac{\operatorname{max}(X_1, \dots, X_n) - d_n}{c_n} \xrightarrow{dist.} H, \;\text{as}\;n\rightarrow\infty.
\end{align*}

\end{definition}

\begin{definition}[Generalized Pareto Distribution]
A random variable is said to have a \textit{Generalized Pareto distribution} (GPD) with shape parameter $c$ and scale parameter $\alpha$ if its cumulative distribution function is given by:
\begin{align*}
    G_{c,\alpha}(x) = \left\{
        \begin{array}{ll}
             1 - \left(1 - \frac{cx}{\alpha}\right)^{1/c}, \quad& c \neq 0 \\
             1 - e^{-x/c}, \quad& c = 0
        \end{array}
            \right.
\end{align*}
The support is $x > 0$ for $c \leq 0$ and $0 < x < \alpha/c$ for $c > 0$.
\end{definition}

\begin{theorem}[Pickands-Balkemade Haan]
Let F be a distribution such that $F \in \text{MDA}(H)$ for some Generalized Extreme Value distribution H, and let $F_u$ be its conditional distribution function. Then,
\begin{align*}
F_u \xrightarrow{dist.} G_{c,\alpha},\quad\text{as}\; u \rightarrow \infty,
\end{align*}
for some $c$ and $\alpha$. In other words, a large family of distributions can have its conditional excess well-approximated by a GPD for sufficiently large threshold $u$.
\label{thm:pickands}
\end{theorem}

With the necessary machinery now in place, let us suppose our retrieval system returns $n$ results with ascending relevance
scores $(s_0, \dots, s_{n-1})$, each with some binary relevance label $l$. Since these scores represent the tail end of relevance scores for the entire, and presumably large, index, we can pretend that they were the result of the following generative process: the scores of observed results with relevance label $l$ are generated by first drawing an i.i.d sample $S_l$ from an unknown distribution $F_l$ and then discarding it if $S_l < u$, for some $u$. In other words, we are effectively drawing from the conditional excess distribution of some unknown $F_l$. Theorem~\ref{thm:pickands}, the \textit{Pickands-Balkemade Haan} theorem~\citep{balkema1974residual}, also referred to as the Second Extreme Value theorem, provides a theoretical justification for using a Generalized Pareto distribution to model the excess distribution of \textit{observed} relevance scores. This is the basis of our proposed method, Surprise scoring.

\begin{algorithm}[!t]
% \small
\caption{Surprise Scoring}
\label{alg:surprise}
\begin{algorithmic}[0]

\Function{GpdRescore}{\text{scores}} \Comment{\parbox[t]{.4\linewidth}{Ascending score list of length $n$}}
\State $i \gets 0, j \gets n$.

\While{\Call{CvmTest}{$\text{scores}[i:j] - \text{scores}[i]$} decreasing}
    \State $j \gets j - 1$.
\EndWhile
\While{\Call{CvmTest}{$\text{scores}[i:j] - \text{scores}[i]$} decreasing}
    \State $i \gets i + 1$
\EndWhile
\State $\text{excess} = \text{scores}[i:] - \text{scores}[i]$.
\State $\text{gpd}$ = \Call{FitGpd}{$\text{excess}$}.
\State $\text{surprise} = \text{zero array of length n}$.
\State $\text{surprise}[i:] = -\log(1 - \text{gpd.cdf}(\text{excess}))$.
\State \textbf{return} \text{surprise}.
\EndFunction

\Function{CvmTest}{excess} \Comment{\parbox[t]{.5\linewidth}{Ascending score excess list of length $m$}}
\State $\text{gpd}$ = \Call{FitGpd}{$\text{excess}$}.
\State $\text{cdf} = \text{gpd.cdf}(\text{excess})$.
\State $\text{linear} = [1 / 2m, 3/2m, ..., (2m-1)/2m]$.
\State $\text{T} = \sum (\text{cdf} - \text{linear})^2 + 1 / (12m)$.
\State \textbf{return} \texttt{T}. \Comment{\parbox[t]{.7\linewidth}{Cramer-von-Mises statistic testing GPD fitness of excess values. Lower is better.}}

\EndFunction

\end{algorithmic}
\end{algorithm}

\begin{figure*}[ht]
\centering
\begin{tabular}{cc}
  \includegraphics[width=0.475\textwidth]{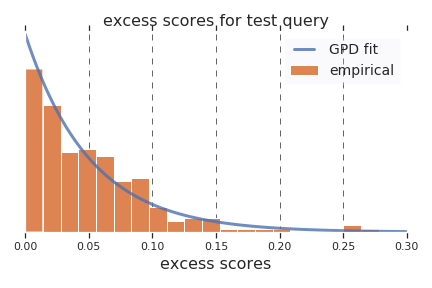} &
  \includegraphics[width=0.475\textwidth]{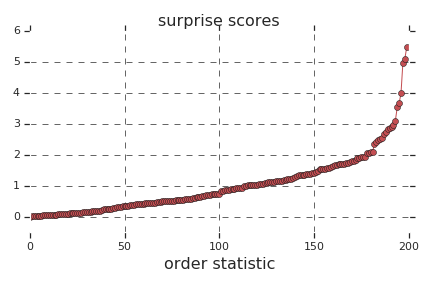}
\end{tabular}
\caption{\textbf{Left:} empirical distribution of excesses for the top-200 results of a sample test query from the Omniglot dataset, along with
the density of the GPD resulting from fitting on the excesses. \textbf{Right:} the Surprise scores for the results by order statistic (relevance increases left to right).}
\label{fig:surprise_example}
\end{figure*}

\section{Surprise Scores}\label{sec:surprisescores}
We now present our proposed method.

\begin{definition}[Surprise Score]
Given Generalized Pareto distribution $G_{c, \alpha}$ with $c \leq 0$ and threshold $u$, define the function $\operatorname{Surprise}_{c, \alpha, u}(s)$ as:
\begin{align*}
\left\{
    \begin{array}{ll}
        -\log(1-G_{c,\alpha}(s-u)), & \quad s \geq u \\
        0, & \quad \text{otherwise}
    \end{array}
\right.
\end{align*}

\end{definition}
The key idea behind our method is to model the conditional excesses of non-relevant
scores using a GPD and assign a new score to each result that captures
how unexpected, or surprising, the result's original score was under the null hypothesis that it came from the non-relevance distribution we fit. Surprise scoring does not change the ranking of the results. Instead, it provides an interpretable and calibrated measure of how relevant each result is compared to non-relevant ones from the result set, much like $p$-values in hypothesis testing. Once each result is rescored, we can truncate the list by picking a threshold that optimizes the target metric of choice. We take as given that the retrieval system has good ranking performance - that is, $S_\text{relevant} > S_\text{non-relevant}$ with high probability - and that a large fraction of the original, non-truncated result list is non-relevant. We can break our algorithm down into four steps.
\begin{enumerate}
    \item We remove scores of results we suspect are relevant (i.e. from $F_\text{relevant}$). We do this by dropping the $n-\hat{j}$ largest scores that hurt the overall model fit. $\hat{j}$ is identified using an iterative greedy procedure that we will describe later.
    \item Now having scores only from $F_\text{non-relevant}$, we proceed to pick the excess cutoff $\hat{u} = s_{\hat{i}}$. To
    find $\hat{i}$, we apply the same method from step 1, but start with the smallest score and work upwards.
    \item We fit a GPD $G_{\hat{c},\hat{\alpha}}$ to the selected samples using maximum likelihood estimation (MLE), constraining $c \leq 0$ so that the GPD has infinite positive support. This is important since we rescore the relevant results we dropped in step 1, whose scores can exceed the ones used in the fit. The constraint has the added benefit that the maximum likelihood estimate is guaranteed to exist and is asymptotically normal and asymptotically efficient \citep{smith1984threshold}.
    \item Rescore each score s:
    \begin{align*}
    s \leftarrow \operatorname{Surprise}_{\hat{c},\hat{\alpha},\hat{u}}(s).
    \end{align*}
\end{enumerate}
\begin{table*}[!ht]
\centering
\begin{tabular}{r|c|cc|cc|}
\multicolumn{1}{l|}{} & Same Document & \multicolumn{1}{l}{Omniglot} & \multicolumn{1}{l|}{} & \multicolumn{1}{l}{News Authorship} & \multicolumn{1}{l|}{} \\ \hline
                 & Accuracy        & \multicolumn{1}{c|}{F1}              & DCG        & \multicolumn{1}{c|}{F1}              & DCG          \\ \hline
Oracle             & 1.0          & \multicolumn{1}{c|}{0.3191}          & 0.6659     & \multicolumn{1}{c|}{0.1122}              & 0.1906          \\
\hline
Global-$k$              & 0.5005          & \multicolumn{1}{c|}{0.2259}          & 0.0     & \multicolumn{1}{c|}{0.0798}              & 0.0          \\
Local-$k$            & 0.4905          & \multicolumn{1}{c|}{0.2265}          & 0.0239    & \multicolumn{1}{c|}{0.0813}              & \textbf{0.0005}           \\
Isotonic    & 0.5723          & \multicolumn{1}{c|}{0.2126}          & 0.0383    & \multicolumn{1}{c|}{0.0689}              & -0.0019           \\
BiCut score only       & 0.3430          & \multicolumn{1}{c|}{0.2175}          & -      & \multicolumn{1}{c|}{0.0789}              & 0.0              \\
BiCut w/ encoding      & 0.4895          & \multicolumn{1}{c|}{0.2225}          & -      & \multicolumn{1}{c|}{0.0733}              & 0.0             \\
\hline
Surprise               & \textbf{0.6112} & \multicolumn{1}{c|}{\textbf{0.2285}} & \textbf{0.0948}   & \multicolumn{1}{c|}{\textbf{0.0814}}              & -0.0213   \\
\hline
\end{tabular}
\caption{Results for the metric learning setting. When matches are present,
they are unusually close to their query and Surprise scores are able to capture this well. Surprise outperforms the baselines on everything except News Authorship DCG.}\label{tab:reuters}
\end{table*}
We now describe the greedy procedure for selecting $\hat{i}$ and $\hat{j}$. The goal is to measure how well each candidate set of excess scores fits a GPD. Various methods for selecting optimal threshold $\hat{u}$ have been suggested (see \cite{langousis2016threshold} for a survey), and although each has its own strengths and weaknesses, the goodness-of-fit tests posed by \cite{choulakian2001goodness} fares well in many practical settings. Given a candidate set of excesses $\{e_0, \ldots, e_m\}$, they suggest:
\begin{enumerate}
    \item Fit a GPD $G_{\tilde{c},\tilde{\alpha}}$ to $\{e_0, \ldots, e_m\}$ using MLE.
    \item Compute the Cramer-von-Mises statistic:
    \begin{displaymath}
    W^2 = \sum_{i=1}^{m} \left(G_{\tilde{c},\tilde{\alpha}}(e_i) - \frac{2i-1}{2m}\right)^2 + \frac{1}{12m}.
    \end{displaymath}
    \item Look up in a pre-computed table the $W^2$ expected under the null hypothesis the excesses come from a GPD with shape $\tilde{c}$. The scale $\tilde{\alpha}$ does not matter.
\end{enumerate}
They suggest setting $u = s_k$ with $k = 0$ initially (so that the initial candidate excesses are $\{0, \dots, s_{n-1}-s_0\}$), and increasing $k$ until
the $p$-value of $W^2$ is within some level of significance. While statistically sound, this method has practical limitations for our setting.
Firstly, it requires computing $W^2$ values for all parameters $c$ and sample size $n$. The authors suggest pre-computing $W^2$ for a finite number of $c$ and then linearly interpolating between known values, but this incurs some error, and since $c$ is unbounded it need not fall near a pre-computed point. Furthermore, in our use case $n$ can
vary from query to query and need not be sufficiently high that falling back to the asymptotic $W^2$ values is justified. Last but not least, their method does not account for the presence of tail, relevant scores that need not come from the hypothesized GPD. Our greedy approach first computes $W^2$ on the entire score list following steps 1 and 2. It then iteratively drops the largest scores and recomputes $W^2$ until the statistic begins to increase. It then applies the same approach to the other side, but dropping the
lowest score each iteration. Algorithm~\ref{alg:surprise} shows our method end to end. Figure~\ref{fig:surprise_example} illustrates the GPD fit and resulting Surprise scores for a sample query.

\section{Experimental Setup}\label{sec:evaluation}
We now describe the experimental setup used to evaluate our proposed Surprise scoring method for the result list truncation task in diverse settings.
\subsection{Datasets}
\subsubsection{Same Document Detection}
We would like to evaluate how well our method is able to decide when to swing or not to swing - that is, when to return results and when not to. To that end, we construct the following problem using the Reuter's corpus of 10,788 news documents (total of 1.3 million words). First, we combine the canonical train and test splits with more than 800 characters and truncate each document to 800 characters. This leaves us with 3,342 documents. We then fit a TF-IDF featurizer using the 1,024 most frequent 1-gram to 6-gram. Each document is subsequently split into a top and bottom half, each consisting of 400 characters, and then featurized. Next, we train a siamese network on the (top, bottom) example pairs using Euclidean distance and contrastive loss with a margin of 1. The encoder is a neural network with 2 1024-unit ReLU layers. We use Adam \citep{kingma2014adam} optimization with default learning rate 0.001. After training the encoder and encoding all examples, we select a random 30\% of the example pairs. Half of the bottom halves are discarded, while the rest are combined with all the examples from the remaining 70\%. We index over this latter set and use the top halves of the 30\% split as a test set. The end result is that half of our test examples will have its matching pair in the index and half will not. Furthermore, since we trained on the combined set, when there is a match, it's particularly close to the query. We want our thresholding methods to return no results when there is no match in the index and to capture it in the result list when there is. Our target metric is accuracy at this task.

\begin{figure*}[!t]
\centering
\begin{tabular}{cc}
  \includegraphics[width=0.475\textwidth]{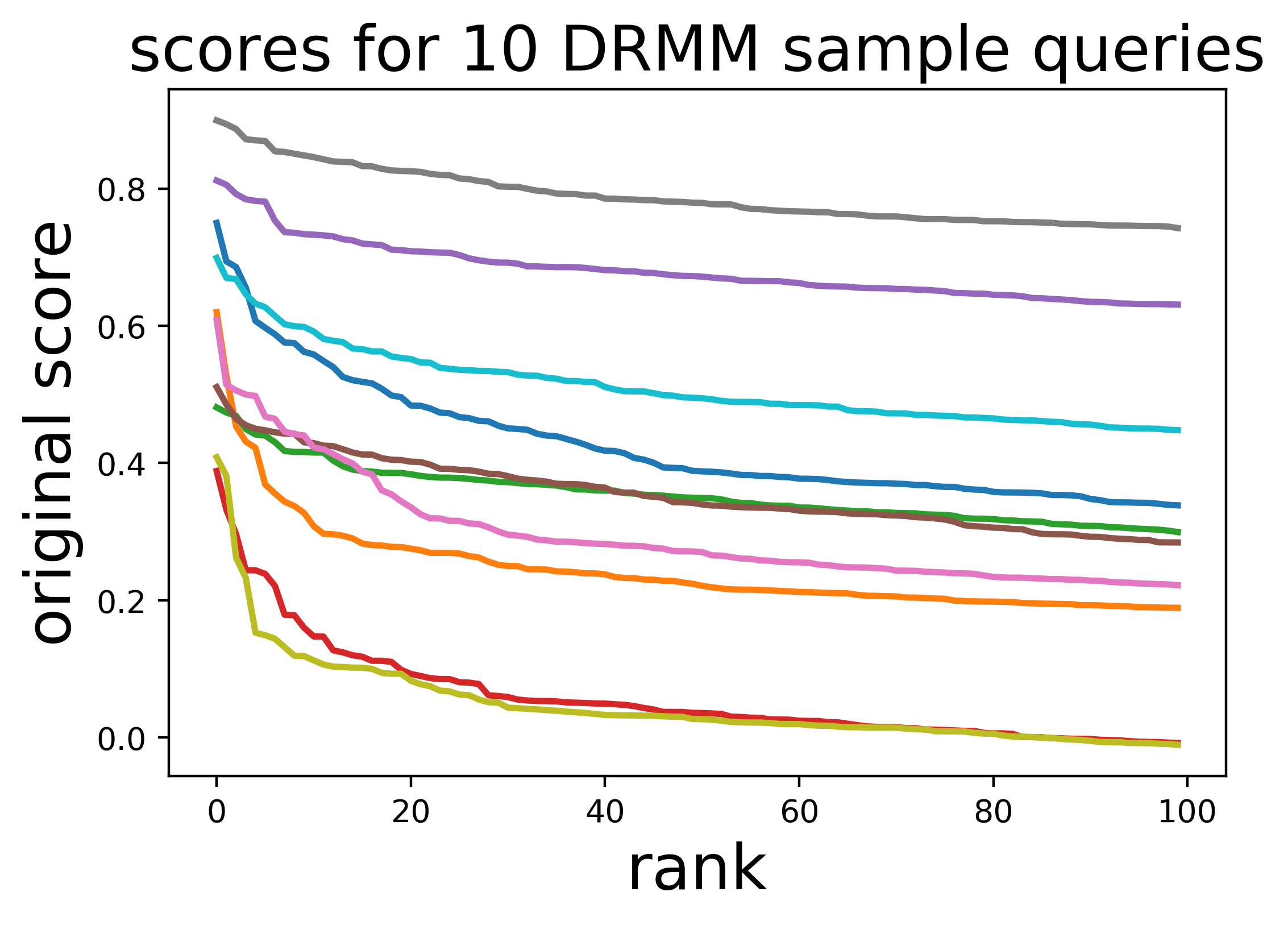} & \includegraphics[width=0.475\textwidth]{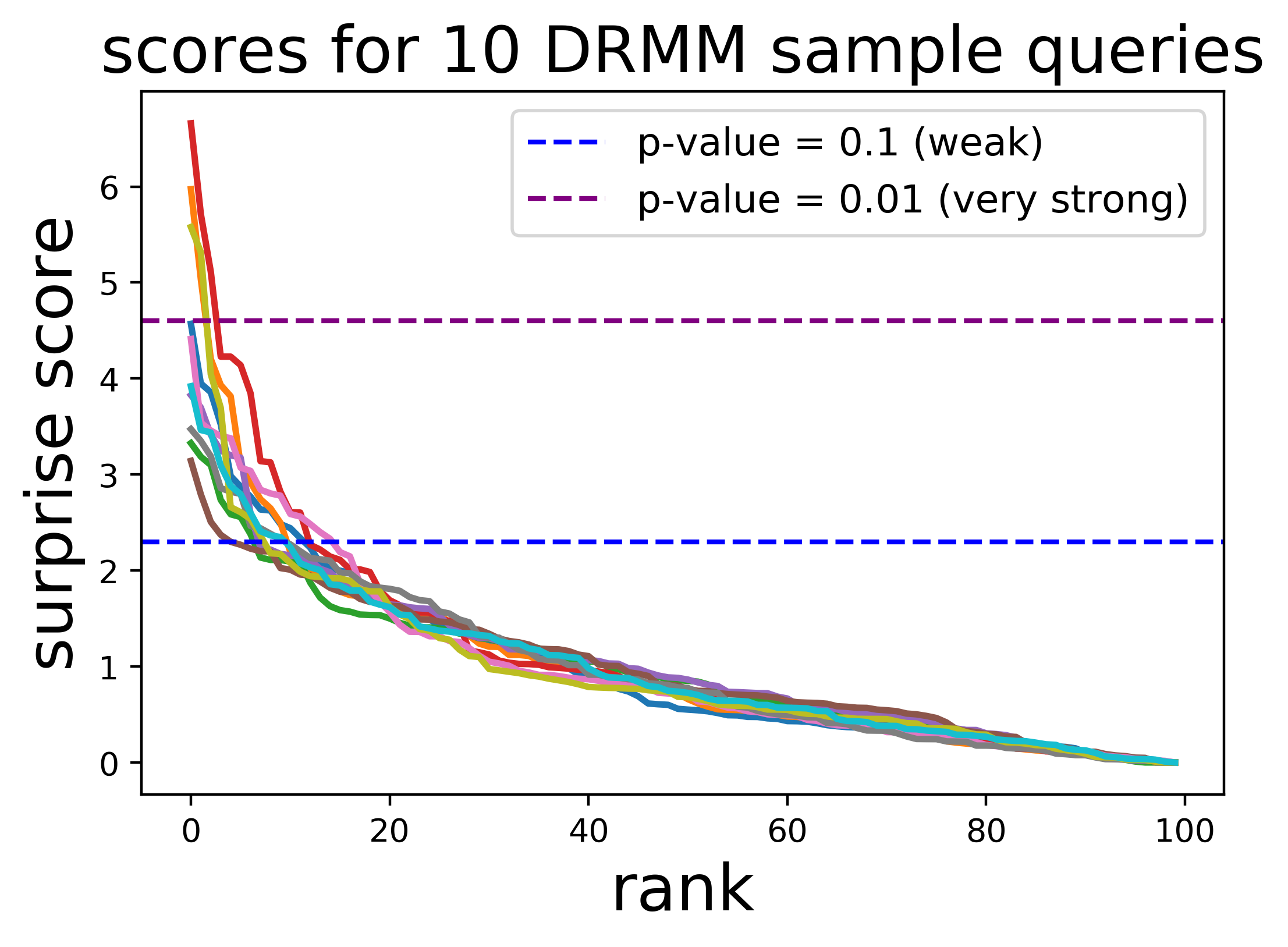} \\
\end{tabular}
\caption{\textbf{Left:} relevance scores for 10 test queries from the Robust04 DRMM test split (relevance decreasing from left to right). \textbf{Right:} Surprise scores for the same 10 queries, along with the thresholds that achieve $p$-values of $0.1$ and $0.01$. These correspond to weak and very strong grounds to reject the null hypothesis respectively, where the null hypothesis here is: ``score is non-relevant''.}
\label{fig:drmm}
\end{figure*}
\subsubsection{News Authorship} 
News articles written by 1,256 authors are partitioned into three sets: train, index, and test. Train (test) consist of 50 (10) articles per author. The index is a subset of train, built as follows: the authors are partitioned into 5 nearly equal sets and for each set a predetermined number of articles are selected for authors in the set. These numbers are 10, 20, 30, 40, 50. Each article has at least 500 words. We featurize each article as a normalized bag of 2,000-most-frequent 4-grams occurring in the training data, and train a siamese network on the character 4-grams features using dot product similarity and batch-based cross-entropy loss. The encoder is a neural network with 1 hidden ReLU layer of 256 units and 1 linear output layer of 128 units.
We generate positive article pairs by randomly sampling 5 articles with the same label and generate negative ones using batch-based negative sampling. More concretely, within each batch of 256 (left, right) example pairs, we have the classification task of predicting each left example's matching right example via softmax over dot product similarities. The loss is cross-entropy, summed over left examples.
We train the siamese network using Adam with learning rate 0.001 for 300 epochs, and then use the trained encoder to encode the index set, chosen in the aforementioned way so that test queries have a varying number of relevant examples.

\subsubsection{Omniglot} 
Omniglot \citep{lake2015human} is a dataset of hand-drawings of 1,623 distinct characters from 50 alphabets, with 20 image examples per character. The dataset is partitioned into train, index and test sets. The train set contains 16 images of each character and the test set contains the remaining 4 images of each character. Index is a subset of train, built the same way as News Authorship: the characters are partitioned into 5 equal sets and for each set a predetermined number of images are selected for characters in the set. These numbers are 4, 7, 10, 13, 16. We train a siamese network using a ResNet-50 \citep{kaiming2015resnet} image encoder and Euclidean distance. We train the encoder using Adam with learning rate 0.001 and then use it to encode the index set.

\subsubsection{Robust04}
To evaluate the effectiveness of our method beyond nearest-neighbor-based retrieval, we use the TREC collection Robust04 from the TREC 2004 Robust Track. It consists of 250 queries over 528k news articles, where each query has 1000 total results and an average of 70 relevant
ones. This is the same dataset used in \cite{lienICTIR19}. We use a random 80/20 train/test that achieves comparable performance to the reported results in \cite{lienICTIR19}. We evaluate the efficacy of our truncation model using two different retrieval approaches - BM25, a traditional tf-idf based model, and DRMM \citep{guo2016deep}, a neural model.

\subsection{Baseline Methods}
We evaluate our method against the following baselines:
\begin{itemize}
\item \textbf{Global-$k$} returns the top-$k$ results, where $k$ is fixed across test queries. The optimal $k$ across training is selected.
\item \textbf{Local-$k$} returns the top-$k$ results, where $k$ is chosen on a per-query basis. The $k$ that achieves the best average metric in the query's neighborhood is chosen.
\item \textbf{Oracle} uses knowledge of each test query's true label to choose the cutoff. This represents an upper bound on the achievable performance.
\item \textbf{Isotonic Regression} gathers the scores and labels of the query's neighbors and learns a isotonic regression model that regresses positive scores to 1 and negative ones to 0. It uses the same neighborhood to pick the optimal threshold and applies the model with this threshold to the query's result scores.

\item \textbf{BiCut}~\citep{lienICTIR19} learns a multi-layer bidirectional LSTM model on the entire training set using the relevance scores and optionally the result's embeddings as inputs. At position $i$ of the result list, the model predicts probability $p_i$ to ``continue'' and probability $1-p_i$ to ``end''. At inference time, the cutoff is made right before the first occurrence of ``end''. The training loss is defined as:
\begin{align*}
\sum_i{w_{y_i} |y_i - p_i|},
\end{align*} where $y$ is the relevance label. $w_y$ is a hyper-parameter which we optimized over the training split following the implementation in the paper.
\end{itemize}

\subsection{Setting}
For every dataset except Robust04, which is not nearest-neighbor based, we fetch the $200$ closest examples from the index set for each query. We then apply each method, which truncate the lists at some point without changing the ordering of the example. We assign a relevance label $\mathrm{rel}$ of $1$ if the example's label matches the query's and $-1$ otherwise.
We report F1 and Discounted Cumulative Gain (DCG) scores in this binary setting, where we define DCG to penalize negative, or non-relevant results as follows:
\begin{align*}
\text{DCG}_p := \sum_{i=1}^p \frac{\mathrm{rel}_i}{\log_2 (i+1)},
\end{align*}
where $p$ is the length of the result list. DCG is not supported by BiCut and is thus omitted from results. No feature normalization was applied to the scores with the exception of max normalization for Isotonic Regression on Robust04 (performance is abysmal without it). In all experiments, the best Surprise score threshold was found by searching the range $[0, 8]$. All methods were implemented using the scipy and scikit-learn python packages.

\begin{table}[!t]
\begin{tabular}{r|cc|cc}
& \multicolumn{2}{c|}{BM25}                           & \multicolumn{2}{c}{DRMM}                          \\ \hline
                      & F1             & DCG             & F1             & DCG            \\ \hline
Oracle                & 0.367          & 1.176           & 0.375          & 1.292          \\
\hline
Global-$k$            & 0.248          & -0.116          & 0.263          & 0.266          \\
Isotonic                 & 0.204          & 0.0               & 0.217          & 0.0              \\
BiCut                 & 0.244          & -               & 0.262          & -              \\
\hline
Surprise                & \textbf{0.251} & \textbf{0.151} & \textbf{0.268} & \textbf{0.295} \\
\hline
\end{tabular}
\caption{Average F1 and DCG performance on Robust04. Surprise achieves competitive results across the board. Since there is no concept of a local neighborhood for this dataset, Local-$k$ is omitted.}\label{tab:robust04_results}
\end{table}
% \vspace{-3mm}

\begin{figure}[!t]
\centering
% \begin{tabular}{c}
  \includegraphics[width=0.475\textwidth]{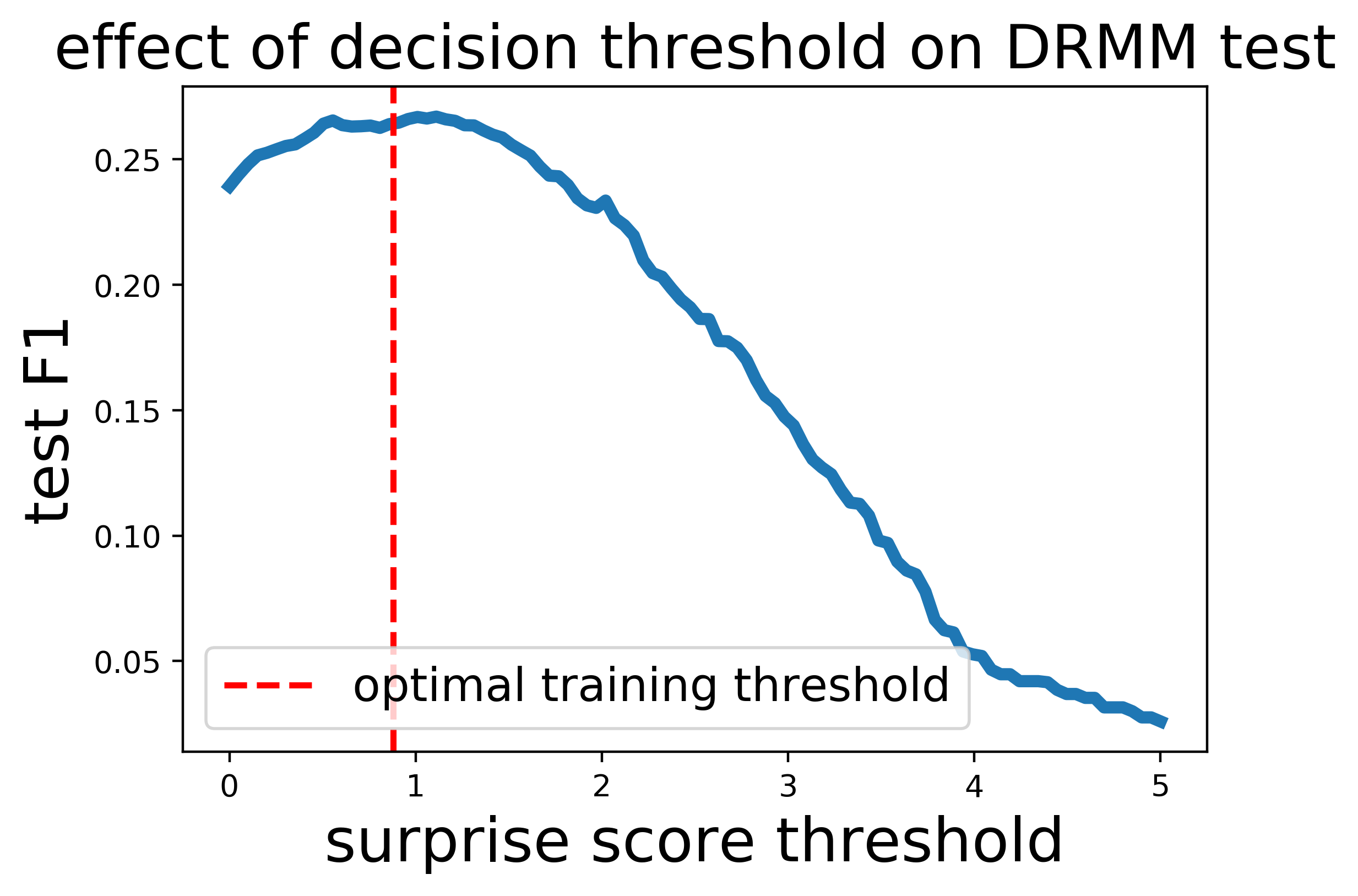} \\
% \end{tabular}
\caption{Test F1 score for different Surprise thresholds on Robust04 DRMM. The optimal test threshold is close to the optimal training one and F1 varies smoothly with threshold.}
\label{fig:drmm_thresholds}
\end{figure}

\section{Experimental Results}
\subsection{Same Document, Omniglot, News Authorship}
Results are shown in Table \ref{tab:reuters}. For the Same Document task, since there are 502 out of 1003 examples with a match in the training set (and in which case the match is always the nearest neighbor), the results for Global-$k$ and Oracle methods perform as expected. Local-$k$ does slightly worse, which is due to the fact that it uses the neighboring training examples to pick $k$ but there is a mismatch between the train and test distributions. In particular, the train set has far more examples with matching pairs than non-matching pairs, while the split is 50/50 for test. Meanwhile, Isotonic Regression performs better than Global-$k$, which is due to the separability of the match and non-match results.
Surprise performs the best as it amplifies this separation. Furthermore, unlike Isotonic Regression which captures the separability over aggregated relevant and non-relevant scores, Surprise scoring does so on an individual query basis. In other words, what matters for Surprise is how the relevant scores are distributed relative to non-relevant ones for \textit{each} query independently, and so it is robust to cases where the variance in distance distribution across examples, even in the local neighborhood, is high.

In a similar vein, Surprise outperforms all other baselines on Omniglot for both F1 and DCG metrics. For the News Authorship data, Surprise fares the best on F1 but struggles with DCG, as does Isotonic Regression. One possible explanation here is that in light of the low Oracle DCG and 0 / near 0 DCG for Local and Global-$k$, for many test examples the optimal strategy is to return no results (this gives a DCG of 0.). This can be challenging for both Surprise and Isotonic Regression since it would require the methods to select a very large decision threshold that rejects all candidate scores.

\subsection{Robust04 Results and Analysis}
Surprise consistently beats the baselines on Robust04. The score distributions from Robust04 are more well-behaved than News Authorship and typical of what one might expect in most retrieval applications. In Figure~\ref{fig:drmm}, we plot the original relevance scores along with Surprise scores for 10 test queries from DRMM. We see that the original scores occupy different ranges but that Surprise calibrates them to be in a consistent range and furthermore amplifies the prominence of highly relevant results. If we consider the hypothesis testing view of the Surprise method -- that is, for every relevance score we test the null hypothesis ``score is non-relevant'' per the non-relevance score distribution we fit using the GPD -- then, quite interestingly, the head of the Surprise score distribution lies in a numerical range that corresponds to $p$-values of $0.1$ to $0.01$. The mapping between $p$-values and Surprise scores is given by:
\begin{align*}
    \text{$p$-value}(s) = \exp(-s),
\end{align*}
where $s$ is the Surprise score. A common evidence scale used in hypothesis testing is to take $0.1$ to be weak evidence to reject the null and $0.01$ to be strong grounds for rejection, i.e. to deem the result not non-relevant. In applications without well-defined target metrics to calibrate against, one can resort to selecting thresholds based on $p$-values. For example, the $p$-value range of $[0.1, 0.01]$ maps to a Surprise score threshold range of $[2.3, 4.6]$.

In Figure~\ref{fig:drmm_thresholds} we plot the test F1 score for different Surprise score thresholds on DRMM. We note that (1) the optimal test threshold is very close to the optimal training one and (2) the performance profile is a smooth, well-behaved function of the threshold.
\section{Conclusion}\label{sec:conclusions}
To summarize, in many practical applications, particularly in search and recommenders, the system returns the best results ranked by some machine-learned relevance score. By design, the ranking and overall quality of these top results are good, but their accompanying scores are often miscalibrated with true relevance. Having a calibrated and understandable ranking score can be useful for users in some applications, enabling them to make statements like ``results with calibrated score value [X] all have about the same relevance to me, regardless of my query''. In addition to being useful to the end user, such calibrated scores are useful programmatically, in the result list truncation task, by allowing the system to truncate the result list using a cutoff on the calibrated scores.

To this end, we proposed \textit{Surprise} score, a statistical method that captures how relevant or surprising a result is by comparing it to non-relevant ones. It does so by leveraging the Generalized Pareto distribution that arises in Extreme Value theory. We apply Surprise to the important task of result list truncation, testing it on not only three distinct nearest-neighbor retrieval-based datasets but also on the benchmark Robust04 dataset. We show that Surprise outperforms both classical and recent baselines in nearly all settings. We illustrate its mechanism of action, its stable performance across thresholds, and show its connection to hypothesis testing and $p$-values.

Potential future work involves testing performance of other distribution-fitting methods and designing experiments to better understand the relationship between retrieval model performance and result list truncation model performance.

% \clearpage
\bibliography{references}

\begin{thebibliography}{}

\bibitem[Abarbanel et~al., 1992]{abarbanel1992statistics}
Abarbanel, H., Koonin, S., Levine, H., MacDonald, G., and Rothaus, O. (1992).
\newblock Statistics of extreme events with application to climate.
\newblock Technical report, MITRE CORP MCLEAN VA JASON PROGRAM OFFICE.

\bibitem[Alvarado et~al., 1998]{alvarado1998modeling}
Alvarado, E., Sandberg, D.~V., and Pickford, S.~G. (1998).
\newblock {\em Modeling large forest fires as extreme events}.
\newblock National Emergency Training Center.

\bibitem[Arampatzis et~al., 2009]{arampatzisSIGIR09}
Arampatzis, A., Kamps, J., and Robertson, S. (2009).
\newblock Where to stop reading a ranked list?: Threshold optimization using
  truncated score distributions.
\newblock In {\em Proceedings of the 32Nd International ACM SIGIR Conference on
  Research and Development in Information Retrieval}, SIGIR '09, pages
  524--531, New York, NY, USA. ACM.

\bibitem[Bahri et~al., 2020]{bahri2020choppy}
Bahri, D., Tay, Y., Zheng, C., Metzler, D., and Tomkins, A. (2020).
\newblock Choppy: Cut transformer for ranked list truncation.
\newblock {\em arXiv preprint arXiv:2004.13012}.

\bibitem[Balkema and De~Haan, 1974]{balkema1974residual}
Balkema, A.~A. and De~Haan, L. (1974).
\newblock Residual life time at great age.
\newblock {\em The Annals of probability}, pages 792--804.

\bibitem[Broder et~al., 2008]{broderCIKM08}
Broder, A., Ciaramita, M., Fontoura, M., Gabrilovich, E., Josifovski, V.,
  Metzler, D., Murdock, V., and Plachouras, V. (2008).
\newblock To swing or not to swing: Learning when (not) to advertise.
\newblock In {\em Proceedings of the 17th ACM Conference on Information and
  Knowledge Management}, CIKM '08, pages 1003--1012, New York, NY, USA. ACM.

\bibitem[Castillo, 2012]{castillo2012extreme}
Castillo, E. (2012).
\newblock {\em Extreme value theory in engineering}.
\newblock Elsevier.

\bibitem[Choulakian and Stephens, 2001]{choulakian2001goodness}
Choulakian, V. and Stephens, M.~A. (2001).
\newblock Goodness-of-fit tests for the generalized pareto distribution.
\newblock {\em Technometrics}, 43(4):478--484.

\bibitem[Cronen-Townsend et~al., 2002]{cronentownsendSIGIR02}
Cronen-Townsend, S., Zhou, Y., and Croft, W.~B. (2002).
\newblock Predicting query performance.
\newblock In {\em Proceedings of the 25th Annual International ACM SIGIR
  Conference on Research and Development in Information Retrieval}, SIGIR '02,
  pages 299--306, New York, NY, USA. ACM.

\bibitem[Culpepper et~al., 2016]{culpepperADCS16}
Culpepper, J.~S., Clarke, C. L.~A., and Lin, J. (2016).
\newblock Dynamic cutoff prediction in multi-stage retrieval systems.
\newblock In {\em Proceedings of the 21st Australasian Document Computing
  Symposium}, ADCS '16, pages 17--24, New York, NY, USA. ACM.

\bibitem[Embrechts et~al., 2013]{embrechts2013modelling}
Embrechts, P., Kl{\"u}ppelberg, C., and Mikosch, T. (2013).
\newblock {\em Modelling extremal events: for insurance and finance},
  volume~33.
\newblock Springer Science \& Business Media.

\bibitem[Guo et~al., 2016]{guo2016deep}
Guo, J., Fan, Y., Ai, Q., and Croft, W.~B. (2016).
\newblock A deep relevance matching model for ad-hoc retrieval.
\newblock In {\em Proceedings of the 25th ACM International on Conference on
  Information and Knowledge Management}, pages 55--64.

\bibitem[Hauff et~al., 2008]{hauffCIKM08}
Hauff, C., Hiemstra, D., and de~Jong, F. (2008).
\newblock A survey of pre-retrieval query performance predictors.
\newblock In {\em Proceedings of the 17th ACM Conference on Information and
  Knowledge Management}, CIKM '08, pages 1419--1420, New York, NY, USA. ACM.

\bibitem[He et~al., 2015]{kaiming2015resnet}
He, K., Zhang, X., Ren, S., and Sun, J. (2015).
\newblock Deep residual learning for image recognition.
\newblock {\em CoRR}, abs/1512.03385.

\bibitem[Kingma and Ba, 2014]{kingma2014adam}
Kingma, D.~P. and Ba, J. (2014).
\newblock Adam: A method for stochastic optimization.
\newblock {\em arXiv preprint arXiv:1412.6980}.

\bibitem[Lake et~al., 2015]{lake2015human}
Lake, B.~M., Salakhutdinov, R., and Tenenbaum, J.~B. (2015).
\newblock Human-level concept learning through probabilistic program induction.
\newblock {\em Science}, 350(6266):1332--1338.

\bibitem[Langousis et~al., 2016]{langousis2016threshold}
Langousis, A., Mamalakis, A., Puliga, M., and Deidda, R. (2016).
\newblock Threshold detection for the generalized pareto distribution: Review
  of representative methods and application to the noaa ncdc daily rainfall
  database.
\newblock {\em Water Resources Research}, 52(4):2659--2681.

\bibitem[Lien et~al., 2019]{lienICTIR19}
Lien, Y.-C., Cohen, D., and Croft, W.~B. (2019).
\newblock An assumption-free approach to the dynamic truncation of ranked
  lists.
\newblock In {\em Proceedings of the 2019 ACM SIGIR International Conference on
  Theory of Information Retrieval}, ICTIR '19, pages 79--82, New York, NY, USA.
  ACM.

\bibitem[Manmatha et~al., 2001]{manmathaSIGIR01}
Manmatha, R., Rath, T., and Feng, F. (2001).
\newblock Modeling score distributions for combining the outputs of search
  engines.
\newblock In {\em Proceedings of the 24th Annual International ACM SIGIR
  Conference on Research and Development in Information Retrieval}, SIGIR '01,
  pages 267--275, New York, NY, USA. ACM.

\bibitem[Montague and Aslam, 2001]{montagueCIKM01}
Montague, M. and Aslam, J.~A. (2001).
\newblock Relevance score normalization for metasearch.
\newblock In {\em Proceedings of the Tenth International Conference on
  Information and Knowledge Management}, CIKM '01, pages 427--433, New York,
  NY, USA. ACM.

\bibitem[Pickands~III et~al., 1975]{pickands1975statistical}
Pickands~III, J. et~al. (1975).
\newblock Statistical inference using extreme order statistics.
\newblock {\em the Annals of Statistics}, 3(1):119--131.

\bibitem[Rudd et~al., 2017]{rudd2017extreme}
Rudd, E.~M., Jain, L.~P., Scheirer, W.~J., and Boult, T.~E. (2017).
\newblock The extreme value machine.
\newblock {\em IEEE transactions on pattern analysis and machine intelligence},
  40(3):762--768.

\bibitem[Shokouhi and Si, 2011]{shokouhiFTIR11}
Shokouhi, M. and Si, L. (2011).
\newblock Federated search.
\newblock {\em Found. Trends Inf. Retr.}, 5(1):1--102.

\bibitem[Smith, 1984]{smith1984threshold}
Smith, R.~L. (1984).
\newblock Threshold methods for sample extremes.
\newblock In {\em Statistical extremes and applications}, pages 621--638.
  Springer.

\bibitem[Tomlinson et~al., 2007]{tomlinsonTREC07}
Tomlinson, S., Oard, D.~W., Baron, J.~R., and Thompson, P. (2007).
\newblock Overview of the trec 2007 legal track.
\newblock In {\em In Proceedings of the 16th Text Retrieval Conference}.

\bibitem[Vignotto and Engelke, 2020]{vignotto2020extreme}
Vignotto, E. and Engelke, S. (2020).
\newblock Extreme value theory for anomaly detection--the gpd classifier.
\newblock {\em Extremes}, pages 1--20.

\bibitem[Wang et~al., 2011a]{wangPAKDD11}
Wang, B., Li, Z., Tang, J., Zhang, K., Chen, S., and Ru, L. (2011a).
\newblock Learning to advertise: How many ads are enough?
\newblock In {\em Proceedings of the 15th Pacific-Asia Conference on Advances
  in Knowledge Discovery and Data Mining - Volume Part II}, PAKDD'11, pages
  506--518, Berlin, Heidelberg. Springer-Verlag.

\bibitem[Wang et~al., 2011b]{wangSIGIR11}
Wang, L., Lin, J., and Metzler, D. (2011b).
\newblock A cascade ranking model for efficient ranked retrieval.
\newblock In {\em Proceedings of the 34th International ACM SIGIR Conference on
  Research and Development in Information Retrieval}, SIGIR '11, pages
  105--114, New York, NY, USA. ACM.

\bibitem[Weng et~al., 2018]{weng2018evaluating}
Weng, T.-W., Zhang, H., Chen, P.-Y., Yi, J., Su, D., Gao, Y., Hsieh, C.-J., and
  Daniel, L. (2018).
\newblock Evaluating the robustness of neural networks: An extreme value theory
  approach.
\newblock {\em arXiv preprint arXiv:1801.10578}.

\bibitem[Zamani et~al., 2018]{zamaniSIGIR18}
Zamani, H., Croft, W.~B., and Culpepper, J.~S. (2018).
\newblock Neural query performance prediction using weak supervision from
  multiple signals.
\newblock In {\em The 41st International ACM SIGIR Conference on Research
  \&\#38; Development in Information Retrieval}, SIGIR '18, pages 105--114, New
  York, NY, USA. ACM.

\bibitem[Zhou and Croft, 2007]{zhouSIGIR07}
Zhou, Y. and Croft, W.~B. (2007).
\newblock Query performance prediction in web search environments.
\newblock In {\em Proceedings of the 30th Annual International ACM SIGIR
  Conference on Research and Development in Information Retrieval}, SIGIR '07,
  pages 543--550, New York, NY, USA. ACM.

\bibitem[Zong and Huang, 2014]{zong2014learning}
Zong, W. and Huang, G.-B. (2014).
\newblock Learning to rank with extreme learning machine.
\newblock {\em Neural processing letters}, 39(2):155--166.

\end{thebibliography}
\end{document}